\documentclass[twocolumn,showpacs,preprintnumbers,amsmath,amssymb]{revtex4}
\usepackage{graphicx}
\usepackage{dcolumn}
\usepackage{bm}
\usepackage[english]{babel}

\begin{document}

\title {Electric fields generated by quantized vortices}

\author {A. S. Rukin}

\author{S. I. Shevchenko}

\affiliation {B. I. Verkin Institute for Low Temperature Physics and
Engineering, Kharkov, Ukraine}

\email{shevchenko@ilt.kharkov.ua}


\begin{abstract}
It is shown that in a magnetic field quantized vortices in a superfluid obtain
a real quantized electric charge concentrated in the vortex core. This charge
is compensated by an opposite surface charge located at a macroscopic distance
from the vortex axis. It is determined that the polarization caused by the
vortex velocity field does not give rise to electric fields outside an infinite
cylinder. Observation of electric fields created by the vortices is possible
only near the end surfaces of the cylinder which must be closed with dielectric
covers to prevent superfluid leaking. Influence of cover properties on the
potential created by the vortex is researched. Potential created by the
vortices on point and ring electrodes are calculated.
\end{abstract}
\maketitle

A series of recent experiments \cite{1}, \cite{2} revealed that flow in
superfluid $^4$He is accompanied by appearance of electric fields in it.
Significant efforts \cite{3} -- \cite{13} were undertaken to understand the
nature of the observed phenomena, but they are unsuccessful up to date.
However, several interesting results were obtained which concern the mechanisms
of polarization of both superfluid and normal systems. For example, in
Melnikovskiy's article \cite{4} it is shown that accelerated motion of a
dielectric medium leads to its polarization proportional to the acceleration.
Natsik \cite{5} -- \cite{7} has studied the peculiarities of polarization of
superfluid systems and, developing Melnikovskiy's ideas, has found that vortex
motion of atoms in a superfluid must lead to their polarization caused by
centrifugal force acting on them. In this case the polarization vector is
directed normally to the velocity of the fluid moving around the vortex line,
so the polarization vectors form a ``hedgehog''. Unfortunately, observation of
electric fields caused by this ``hedgehog'' is rather difficult due to their
rapid decrease with distance from the vortex line (reverse proportional to the
cube of the distance). The authors \cite{14A} -- \cite{15} found that in a
magnetic field the vortex line acquires a real electric charge whose magnitude
is proportional to the vortex circulation and is quantized in the same way as
the circulation. The compensating charge of the opposite sign appears on the
surface of the system. Macroscopic spatial separation of the vortex charge and
the compensating one allows to measure the electric field caused by these
charges and to observe the motion of vortices in the superfluid. In this message
we discuss the significant aspects of measuring the electric fields generated
by the vortices.

First we recall the arguments of the articles \cite{14A} -- \cite{15}. In a
magnetic field $\bf H$ a rarefied medium moving with velocity $\bf v$ acquires
polarization
\begin{equation}\label{1}
{\bf{P}} = n\alpha \frac{{{\bf{v}} \times {\bf{H}}}}{c}.
\end{equation}
Here $\alpha$ is the atom polarizability, $n$ is the medium density, $c$ is the
velocity of light. It follows from the common expression for the electric
induction of the moving medium
\begin{equation}\label{2}
{\bf{D}} \equiv {\bf{E}} + 4\pi {\bf{P}} = \epsilon {\bf{E}} +
\frac{{\epsilon \mu  - 1}}{c}{\bf{v}} \times {\bf{H}},
\end{equation}
where $\mu$ is the magnetic permeability, that the expression (\ref{1}) for the
dipole moment $\bf P$ is valid when $\mu=1$ and $\epsilon=1+4\pi n\alpha$.

In a superfluid there exists a ``characteristic configuration'' of the velocity
field $\bf v$ caused by a vortex,
\begin{equation} \label{3}
{{\bf{v}}_s} = \frac{\hbar }{M}\nabla \phi.
\end{equation}
Here $\phi$ is the phase of the order parameter. In the case of a rectilinear
vortex in a uniform magnetic field this velocity field generates a polarization
charge
\begin{multline} \label{4}
{\rho _{pol}} \equiv  - {\mathop{\rm div}\nolimits} {\bf{P}} =  - \frac{\alpha
}{c}{n_s}{H_z}{({\mathop{\rm rot}\nolimits} {{\bf{v}}_s})_z} =   \\
 =  - \frac{{\alpha {n_s}}}{c}{H_z}\frac{{2\pi \hbar }}{M}s{\delta _2}({\bf{r}} -
 {{\bf{r}}_v}),
\end{multline}
where $s=\pm 1$, $\delta_2(\bf r-\bf r_v)$ is the 2D $\delta$-function in the
plane perpendicular to the vortex axis. We denote this axis by $z$ and assume
that the magnetic field is directed along this axis. Thus, a polarization
charge appears on the axis, and its density per unit vortex length equals to
\begin{equation} \label{5}
{q_v} =  - \alpha {n_s}H\frac{{2\pi \hbar }}{M}s.
\end{equation}
The compensating charge appears on the surface, and its density equals to the
normal component of the polarization vector $\bf P$
\begin{equation} \label{6}
{\sigma _s} = \alpha {n_s}\frac{{{{\left[ {{{\bf{v}}_s}({\bf{r}}) \times
{\bf{H}}} \right]}_n}}}{c}.
\end{equation}
The total surface charge equals to
\begin{equation} \label{7}
\int {{\sigma _s}dS = \frac{{\alpha {n_s}\hbar }}{{Mc}}\int {\left( {\nabla
\phi  \times {\bf{H}}} \right) \cdot d{\bf{S}}} }.
\end{equation}
Now we confine ourselves to the case when the vessel with superfluid $^4$He has
the form of a cylinder and the vortices are rectilinear and directed parallel
to the cylinder axis. In this case the surface element $d\bf S$ can be replaced
with $L_z d\bf l$, where $L_z$ is the vessel length and $d\bf l$ is the vector
in the $(x,y)$ plane. The total surface charge per unit length along the $z$
axis equals to
\begin{multline} \label{8}
\frac{{\alpha {n_s}\hbar }}{{Mc}}\int {{{\left( {\nabla \phi  \times {\bf{H}}}
\right)}_n} \cdot d{{\bf{l}}_n}}  =  \\
= \frac{{\alpha {n_s}\hbar
}}{{Mc}}{H_z}\int {\nabla \phi  \cdot } d{{\bf{l}}_\tau } = \frac{{\alpha
{n_s}\hbar }}{{Mc}}{H_z} \cdot 2\pi s.
\end{multline}
We have taken into account that the phase incursion around the vortex equals to
$2 \pi s$. Comparing this result with the result (\ref{4}) shows that the
electric neutrality condition is satisfied and the total polarization charge
equals to zero.

Here we must note that the distance between the vortex and the vessel wall is
in general macroscopically large. However, observation of the electric fields
caused by the vortex is a complicated experimental task. The fact is that in
the case of an infinite cylinder the electric field generated by the dipole
moment $\bf P$ from (\ref{1}) is nonzero only inside the cylinder. Really, in the
absence of external charges the electric induction $\bf D$ satisfies the
equation
\begin{equation} \label{9}
{\mathop{\rm div}} {\bf{D}} = 0.
\end{equation}
On the other hand, taking into account that in a stationary case ${\mathop{\rm
rot}} {\bf{E}} = 0$, we obtain for the components of $\mathop {\rm rot} \bf D$
\begin{gather}
{{\mathop{\rm rot}\nolimits} _z}{\bf{D}} =  - 4\pi \frac{{\alpha
{n_s}}}{c}{H_z}\left( {\frac{{\partial {v_{sx}}}}{{\partial x}} +
\frac{{\partial {v_{sy}}}}{{\partial y}}} \right);\label{10}\\
{{\mathop{\rm rot}\nolimits} _x}{\bf{D}} = 4\pi \frac{{\alpha
{n_s}}}{c}{H_z}\frac{{\partial {v_{sx}}}}{{\partial z}};\label{11}\\
{{\mathop{\rm rot}\nolimits} _y}{\bf{D}} = 4\pi \frac{{\alpha
{n_s}}}{c}{H_z}\frac{{\partial {v_{sy}}}}{{\partial z}}.\label{12}
\end{gather}
In an infinite cylinder the velocity field of the vortex does not depend on
$z$. In this case ${{\mathop{\rm rot}} _x}{\bf{D}} = {{\mathop{\rm rot}}
_y}{\bf{D}} = 0$. The $z$ component ${{\mathop{\rm rot}} _z}{\bf{D}}$ also
equals to zero because in the stationary case ${\mathop{\rm div}} {\bf{v}} =
0$. Therefore in this case the vector $\bf D$ is constant which is equal to
zero, as far as at infinity outside the cylinder the polarization $\bf P$ is
zero and the electric field $\bf E$ is also zero. As the result, inside the
infinite cylinder the electric field is
\begin{equation} \label{13}
\bf E =-4\pi \bf P,
\end{equation}
and outside the cylinder it equals to zero. Observation of the electric field
is possible only for a finite sized cylinder in the vicinity of its end
surfaces.

Before writing down the corresponding expression, we remind that in superfluid
$^4$He a phenomenon called ``film flow'' takes place. Superfluid helium filling
a laboratory glass begins to climb up its walls and flow out over the edge.
Therefore experiments on the behavior of rectilinear vortices in a cylindrical
vessel must be carried out in a closed system, i.~e. the end of the cylinder with helium
must be closed with a cover. We will assume this condition to be satisfied and
consider separately two cases: a) the measuring electrode (which detects the
potential caused by the vortex) is outside the cover and outside the cylinder;
b) the electrode is inside the cover. We assume that the dielectric constant of
the cover is $\epsilon$ and its thickness is $d$. Presence of a dielectric near
a system of charges leads to change of electric field in the space. We will
consider the influence of a dielectric cover (plate) on the electric field of a
point charge.

The problem of a point charge $e$ near a dielectric plate of finite thickness
$d$ has an exact solution \cite{16}. In the case a) the potential outside the
plate on the other side of the charge equals to
\begin{equation} \label{14}
\phi (\rho ,z) = \int\limits_0^\infty  {e\frac{{1 - {\beta ^2}}}{{1 - {\beta
^2}\exp ( - 2kd)}}{J_0}(k\rho )\exp ( - kz)dk.}
\end{equation}
Here $\beta  = \frac{{\epsilon  - 1}}{{\epsilon  + 1}}$, $J_0$ is the Bessel
function. It is possible to calculate the integral when $z\gg d$. In this case
due to the factor $\exp(-kz)$ significant values of $k$ are of the order of
$1/z$. This fact allows to use the expansion $\exp ( - 2kd) \approx 1 - 2kd$
and to obtain after a simple integration
\begin{equation} \label{15}
\phi (\rho ,z) = \frac{e}{{\sqrt {{z^2} + {\rho ^2}} }} -
\frac{{e{{(\epsilon  - 1)}^2}zd}}{{\epsilon {{({z^2} + {\rho
^2})}^{3/2}}}}.
\end{equation}
During the calculation it was also assumed $\frac{{{{(\epsilon  -
1)}^2}}}{\epsilon }\frac{d}{z} \ll 1$. It follows from (\ref{15}) that with
the assumptions made here the electric field behind the plate coincides in the
zero approximation with the electric field of a point charge in vacuum. The
correction increases with increasing $\epsilon$.

In the case b) the potential inside the dielectric plate is given by an
expression
\begin{multline} \label{16}
\phi (\rho ,z) = \int\limits_0^\infty  {e\frac{{(1 - \beta )}}{{1 - {\beta
^2}\exp ( - 2kd)}}}\times\\ {\times\left\{ {\exp ( - kz) + \exp \left[ { - 2k(d
+ z') + kz} \right]} \right\}dk},
\end{multline}
where $z'$ is the distance from the charge to the surface of the plate which is
in contact with helium. Integrals in (\ref{16}) can be calculated when $z\ll
d$. In this case values $k \sim 1/z$ are significant and $\exp ( - 2kd) \ll 1$.
Expanding the denominator in this small addition, we obtain after integrating
\begin{multline} \label{17}
\phi  = \frac{2}{{\epsilon  + 1}}\frac{e}{{\sqrt {{z^2} + {\rho ^2}} }} +\\+
\frac{{2e(\epsilon  - 1)}}{{{{(\epsilon  + 1)}^2}}}{\left( {{\rho ^2} +
{{(z + 2z' + 2d)}^2}} \right)^{ - 1/2}} +\\+ \frac{{2e{{(\epsilon  -
1)}^2}}}{{{{(\epsilon  + 1)}^3}}}{\left( {{\rho ^2} + {{(z + 2d)}^2}}
\right)^{ - 1/2}}.
\end{multline}
This result corresponds (in the zero approximation) with the result which can
be obtained for a dielectric semi-infinite space using the electric image
method (see e.~g. \cite{17}): the potential inside the dielectric is the
potential of the point charge in vacuum multiplied by a constant coefficient
$\frac 2{\epsilon+1}$.

Expressions (\ref{15}), (\ref{17}) show that calculating the potential of a
system of charges in both cases (neglecting the corrections) is reduced to
finding the potential of the same system of charges in vacuum. If the electrode
is mounted inside the cover, the result is multiplied by $\frac
2{\epsilon+1}$.

First we find the potential created by a vortex on the end surface of the
cylinder with helium, more exactly, at $z=+0$.
\begin{multline} \label{18}
\phi (\rho , + 0) = \int {\frac{{{\bf{P}}({\bf{r'}}) \cdot ({\bf{r}} - {\bf{r'}})}}{{{{\left| {{\bf{r}} - {\bf{r'}}} \right|}^3}}}d{\bf{r'}}}  =\\= \int\limits_{ - \infty }^0 {\int {\frac{{{\bf{P}}({\bf{\rho '}}) \cdot ({\bf{\rho }} - {\bf{\rho '}})}}{{{{\left[ {{{({\bf{\rho }} - {\bf{\rho '}})}^2} + {{z'}^2}} \right]}^{3/2}}}}{d^2}\rho 'dz'} }  = \\
 = \frac{1}{2}\int\limits_{ - \infty }^\infty  {\int {\frac{{{\bf{P}}({\bf{\rho '}}) \cdot ({\bf{\rho }} - {\bf{\rho '}})}}{{{{\left[ {{{({\bf{\rho }} - {\bf{\rho '}})}^2} + {{z'}^2}} \right]}^{3/2}}}}{d^2}\rho 'dz'} } .
\end{multline}
Thus, the problem is reduced to finding the potential of a vortex in an
infinite cylinder. However, instead of finding $\phi (\rho , + 0)$ by
calculating the integrals in (\ref{18}), it is much simpler to obtain the
electric field created by the vortex. Due to (\ref{13}) the field is $\bf
E=-4\pi\bf P$. We limit ourselves with a circular cylinder of radius $R$ and
assume that the length of the cylinder $L$ and the radius $R$ satisfy the
inequality $L \gg R$. The velocity field of a rectilinear vortex is a sum of
the velocity field of the vortex line with circulation $\oint
{{{\bf{v}}_s}d{\bf{l}} = 2\pi \kappa } $ located at the point ${{\bf{r}}_0} =
({r_0},{\theta _0})$ and its ``image'' with circulation $-2\pi\kappa$ located at
${{\bf{r}}_0'} = ({R^2}/{r_0},{\theta _0})$:
\begin{equation} \label{19}
{{\bf{v}}_s} = \frac{\kappa }{{\left| {{\bf{r}} - {{\bf{r}}_0}}
\right|}}{{\bf{\hat e}}_\theta } - \frac{\kappa }{{\left| {{\bf{r}} -
{{{\bf{r}}}_0'}} \right|}}{{\bf{\hat e}}_{\theta '}},
\end{equation}
where ${\bf{\hat e}}_\theta$ and ${\bf{\hat e}}_{\theta'}$ are unit vectors
tangent to circles with centers on the axis of the vortex line and its image
correspondingly. In (\ref{19}) -- (\ref{24}) $\bf r$ and ${\bf r}_0$ are 2D
vectors. In the Cartesian coordinates this expression equals to
\begin{equation} \label{20}
{{\bf{v}}_s}({\bf{r}}) = \kappa {\bf{\hat z}} \times \nabla \ln \frac{{\left|
{{\bf{r}} - {{\bf{r}}_0}} \right|}}{{\left| {{\bf{r}} - {{{\bf{r}}}_0'}}
\right|}}.
\end{equation}
Substituting this result to (\ref{1}), we obtain the polarization $\bf P$ and
the electric field $\bf E$ inside the cylinder
\begin{equation} \label{21}
{\bf{E}} =  - \frac{{4\pi \alpha {n_s}\kappa {H_z}}}{c}\nabla \ln \frac{{\left|
{{\bf{r}} - {{\bf{r}}_0}} \right|}}{{\left| {{\bf{r}} - {{{\bf{r}}}_0'}}
\right|}}.
\end{equation}
Taking into account that $\bf E=-\nabla\phi$ and using (\ref{21}), we find
\begin{equation} \label{22}
\phi  = 4\pi \frac{{\alpha {n_s}\kappa {H_z}}}{c}\ln \frac{{\left| {{\bf{r}} -
{{\bf{r}}_0}} \right|}}{{\left| {{\bf{r}} - {{{\bf{r}}}_0'}} \right|}} + C.
\end{equation}
The integration constant $C$ must be chosen to satisfy the condition: the total
potential created by the vortex and its ``image'' on the side surface of the
cylinder equals to zero. This boundary condition is a consequence of the fact
that outside the cylinder the electric field and its potential are zero, and as
far as the potential must be a continuous function of coordinates (otherwise,
infinite electric fields would appear at its discontinuity points), on the
surface of the cylinder also $\phi\equiv 0$. Assuming $\phi=0$ at ${\bf r}={\bf
R}$, where $\bf R$ is an arbitrary radius vector in the $(x,y)$ plane drawn
from the axis to the side surface of the cylinder, we find
\begin{equation} \label{23}
C =  - 4\pi \frac{{\alpha {n_s}\kappa {H_z}}}{c}\ln \frac{{\left| {{\bf{R}} -
{{\bf{r}}_0}} \right|}}{{\left| {{\bf{R}} - {{{\bf{r}}}_0'}} \right|}} = 4\pi
\frac{{\alpha {n_s}\kappa {H_z}}}{c}\ln \frac{R}{r_0}.
\end{equation}

The electric potential created in a semi-infinite cylinder on the internal
surface of its dielectric cover equals to
\begin{equation} \label{24}
\phi  = \frac{{4\pi }}{{\epsilon  + 1}}\frac{{\alpha {n_s}\kappa
{H_z}}}{c}\ln\left[ \frac{{\left| {{\bf{r}} - {{\bf{r}}_0}} \right|}}{{\left|
{{\bf{r}} - {{{\bf{r}}}_0'}} \right|}}\frac{R}{{{r_0}}}\right].
\end{equation}
This potential depends on the position of the vortex relative to the electrode.
Although the change of the potential on changing the vortex position is weak,
it can be measured. For liquid helium (density $n=2\cdot 10^{22}
\mathrm{cm}^{-3}$, polarizability $\alpha=2\cdot 10^{-25} \mathrm{cm}^3$) in a
magnetic field $10^5 \mathrm{Gs}$ the potential has an order of magnitude of
$10^{-8} \mathrm V$.

Measuring the potential with a point electrode is not always a simple
experimental problem, especially in a rotating vessel with helium. Therefore we
consider also a situation that takes place if the electrode is a thin metallic
ring mounted into the cover in the $(x,y)$ plane. We assume that the center of
the ring is on the cylinder axis and its radius is $r$. The potential at an
arbitrary point $\bf r$ on the ring can be expressed in terms of the charges on
the ring $q(\bf r)$, on the side surface of the cylinder $Q(\bf R)$ and the
vortex charge $q_v$:
\begin{equation} \label{25}
\phi ({\bf{r}},0) = \int {\frac{{q({\bf{r'}})}}{{\left| {{\bf{r}} - {\bf{r'}}}
\right|}}d{\bf{r'}}}  + \int {\frac{{Q({\bf{R}})}}{{\left| {{\bf{r}} -
{\bf{R}}} \right|}}d{\bf{R}}}  + \int {\frac{{{q_v}({{\bf{r}}_v})}}{{\left|
{{\bf{r}} - {{\bf{r}}_v}} \right|}}d{{\bf{r}}_v}}.
\end{equation}
Here $q(\bf r)$ is the charge induced on the ring electrode by the vortex
charge and the surface charge of the side surface $Q(\bf R)=P_n(\bf R)$, where
$P_n(\bf R)$ is the normal component of the polarization vector,
${q_v}({\bf{r}}) =  - {\mathop{\rm div}} {\bf{P}}({\bf{r}})$ is the density of
the vortex charge.

Firstly it seems that we must know the charge distribution $q(\bf r)$ in order
to find the potential on the ring. But it turns out that in the case of a
conducting ring its equipotentiality allows to avoid the calculation of $q(\bf
r)$. Indeed, using polar coordinates while integrating (\ref{25}), one can
write down this expression as
\begin{multline} \label{26}
\phi  = \int {\frac{{q(\theta ')d\theta '}}{{\sqrt {2\left[ {1 - \cos (\theta -
\theta ')} \right]} }}}  +\\+ \int {\frac{{Q(\theta ')Rd\theta 'dZ}}{{\sqrt
{{R^2} + {r^2} - 2Rr\cos (\theta ' - \theta ) + {Z^2}} }}}  +\\+ \int
{\frac{{{q_v}dz}}{{\sqrt {r_0^2 + {r^2} - 2{r_0}r\cos \theta  + {z^2}} }}}.
\end{multline}
Now we integrate both sides of this equation by $\theta$ along the whole ring.
In the right side we replace the variables $\theta-\theta'=\alpha$, and after
this substitution the angular variables separate themselves. Taking into
account that $\phi$ is independent of $\theta$, we obtain
\begin{multline} \label{27}
\phi \int {d\theta  = \int {q(\theta ')d\theta '} } \int {\frac{{d\alpha }}{{\sqrt {2(1 - \cos \alpha )} }}}  +\\+ \int {Q(\theta ')d\theta '} R\int {\frac{{d\alpha dz}}{{\sqrt {{r^2} - 2rR\cos \alpha  + {R^2} + {z^2}} }}}  + \\
 + {q_v}\int {\frac{{d\theta dz}}{{\sqrt {{r^2} - 2r{r_0}\cos \theta  + r_0^2 + {z^2}} }}} .
\end{multline}
As far as the whole ring electrode remains electrically neutral, the total
charge induced on it is zero, i. e. $\int {q(\theta ')d\theta '}  = 0$, so, the
unknown charge distribution $q(\theta)$ falls out of the problem.

On the other hand, the condition of electric neutrality of helium yields $\int
{Q(\theta )Rd\theta  = -{q_v}} $. As the result, after integrating by $z$ (from
$-\infty$ to 0) in the two remaining integrals in (\ref{27}) we obtain
\begin{equation}\label{28}
2\pi \phi  = \frac{{{q_v}}}{2}\int\limits_0^{2\pi } {\ln \frac{{1 -
2\frac{{{r_0}}}{r}\cos \theta  + {{\left( {\frac{{{r_0}}}{r}} \right)}^2}}}{{1
- 2\frac{R}{r}\cos \theta  + {{\left( {\frac{R}{r}} \right)}^2}}}} d\theta.
\end{equation}
In order to calculate the integrals in this expression
\begin{equation} \label{29}
I(a) = \int\limits_0^{2\pi } {\ln \left( {1 - 2a\cos \theta  + {a^2}}
\right)d\theta }
\end{equation}
we use the following method. Differentiating both sides of the equation by $a$,
we obtain
\begin{equation} \label{30}
\frac{{dI}}{{da}} = \int\limits_0^{2\pi } {\frac{{ - 2\cos \theta  + 2a}}{{1 -
2a\cos \theta  + {a^2}}}d\theta }.
\end{equation}
Now it is easy to perform the integration by $\theta$.
\begin{equation} \label{31}
\frac{{dI}}{{da}} = \frac{\pi }{a}\left( {1 - a\mathrm{sign}(1 - {a^2})}
\right).
\end{equation}
Integrating this equation by $a$ from 0 to $a$, we find
\begin{equation} \label{32}
I(a) = \left\{ {\begin{array}{*{20}{l}}
{0,}&{a \le 1;}\\
{2\ln a,}&{a > 1.}
\end{array}} \right.
\end{equation}
Returning to (\ref{28}) and taking into account this result, we find the
potential of the ring with radius $r$
\begin{equation} \label{33}
\phi  = {q_v}\left\{ {\begin{array}{*{20}{l}}
{\ln \frac{R}{{{r_0}}},}&{{r_0} \ge r;}\\
{\ln \frac{R}{r},}&{{r_0} < r.}
\end{array}} \right.
\end{equation}
We see that the potential $\phi$ depends on the vortex coordinate $r_0$ only if
the vortex is outside the measuring ring electrode. If the vortex is inside the
ring, the potential depends only on the electrode radius $r$. For a system of
vortices the total potential is a sum of potentials created by all vortices.

The information about the distribution of vortices in the vessel with helium
will be more complete if we use two ring shaped measuring electrodes with
radiuses $R_1$ and $R_2$ ($R_2>R_1$). Analysis similar to the one performed
above shows that the potential difference between the electrodes equals to
\begin{equation} \label{34}
\delta \phi  \equiv {\phi _1} - {\phi _2} = {q_v}\left\{
{\begin{array}{*{20}{l}}
{0,}&{{r_0} > {R_2};}\\
{\ln ({R_2}/{r_0}),}&{{R_2} \ge {r_0} > {R_1};}\\
{\ln ({R_2}/{R_1}),}&{{R_1} \ge {r_0}.}
\end{array}} \right.
\end{equation}
If the vortex is outside both electrodes the potential difference between the
rings is absent. When the vortex passes the external ring (it is assumed that
the vortex moves from the side surface to the axis), the potential difference
increases and becomes constant and independent of the vortex position when the
vortex passes inside the internal ring electrode. This passage of the vortex
into the rings is determined the more exactly, the less is the distance between
the rings. Thus, installing ring shaped electrodes with various radiuses, one
can register the quantity, positions and motion of vortices in the vessel.

It is useful to write down the expression for the average potential $\phi(r)$
which is measured by a ring electrode for a system of vortices with identical
circulations (and therefore with identical charges $q_v$), distributed
homogeneously in the vessel. The potential in this case equals to
\begin{equation} \label{35}
\phi (r) = {q_v}\sum\limits_{{\bf{r}}_i} {\left\{ {\Theta \biggl(
{\frac{r}{r_i} - 1} \biggr)\ln \frac{R}{r} + \Theta \biggl( {\frac{r_i}{r} - 1}
\biggr)\ln {\frac{R}{r_i}} } \right\}}.
\end{equation}
Here $\Theta(x)$ is the Heaviside step function: $\Theta(x<0)=0$ and
$\Theta(x\ge 0)=1$. Introducing the density of vortices $n_v$ and replacing the
summation by ${\bf r}_i$ with 2D integration, we obtain
\begin{multline} \label{36}
\phi (r) =\\= 2\pi {q_v}{n_v}\left[ {\int\limits_0^r {\left( {\ln \frac{R}{r}}
\right){r_1}} d{r_1} + \int\limits_r^R {\left( {\ln \frac{R}{{{r_1}}}}
\right){r_1}} d{r_1}} \right] =\\= \pi {q_v}{n_v}({R^2} - {r^2}).
\end{multline}
It is obvious that the potential reaches its maximum value in the center of the
vessel, i. e. at $r=0$. Homogeneously distributed vortices form a lattice (most
probably a triangular one). The potential $\phi$ obtained above is the value of
potential averaged by all possible positions of a lattice with given density
$n_v$.

As it is well known, rectilinear vortices are formed in a cylindrical vessel
when it rotates with angular velocity $\Omega$ greater than the critical
velocity ${\Omega _c} = (\hbar /M{R^2})\ln (R/\xi )$, where $\xi$ is the
coherence length. If $\Omega \gg \Omega_c$, the quantity of vortex lines will
be very large, and in the $\Omega \to \infty$ limit they imitate rotation of
the superfluid component as a whole with angular velocity $\Omega$. The
condition $\mathrm{rot} {\bf v}_s=2\Omega$ gives the density of the vortices
\begin{equation}\label{37}
n_v=\frac{m\Omega}{\pi\hbar}.
\end{equation}
Substituting this expression for $n_v$ into (\ref{36}) and replacing $q_v$ with
its value from (\ref{5}), we obtain the result
\begin{equation} \label{38}
\phi (r) = \frac{{2\pi \alpha {n_s}H}}{c}\Omega ({R^2} - {r^2}).
\end{equation}
We can get exactly the same result if we assume that a liquid with density
$\rho_s$ performs solid rotation with velocity ${\bf v}_s={\bf
\Omega}\times{\bf r}$, find the polarization (\ref{1}) for this velocity field,
the corresponding electric field ${\bf E}=-4\pi{\bf P}$ and integrate this
field from $r$ to $R$. Thus, the theory developed above, which operates with
separate quantized vortices, gives a correct description of the system's
behavior in the continuous medium limit.

Until now we considered only the superfluid component. But at nonzero
temperatures there always exists the normal component. In a vessel rotating
with angular velocity $\Omega$ the normal component performs solid rotation with velocity
${\bf{v}} = {\bf{\Omega }} \times {\bf{r}}$. Calculating the polarization
(\ref{1}) for this velocity field, we find the addition to the electric field
and its potential $\phi(\bf r)$ caused by the normal component. The addition to
the potential is obtained from (\ref{38}) by replacing the superfluid density
$n_s$ with the normal density $n_n$. Therefore, to observe the effects caused
by separate vortices, one must carry out the experiments at temperatures
sufficiently low to make the normal density much lower than the superfluid
density. For He II this condition is satisfied below 1 K.

It is more convenient to measure not the static potential, but its variation in
time. When a vortex passes inside (or outside) a pair of measuring ring
electrodes, the potential difference between them undergoes a jump. If more
than one vortex passes, the change of the potential is correspondingly
multiplied by the quantity of the vortices.

Let us briefly consider the behavior of vortices in a rotating vessel. Using the known
expression for the free energy of a vortex in a rotating vessel \cite{18}, one
can obtain that at angular velocity $\omega  > {\omega _1} \equiv \kappa
/{R^2}$ the energy has a maximum at a distance ${x_0} = \sqrt {{R^2} - \kappa
/\omega } $ from the axis and a minimum on the axis. Vortices formed near the
walls of the cylinder try to overcome the potential barrier and accumulate on
its external side. Therefore it is reasonable to place the measuring
electrodes with radius $x_0$.

To observe the motion of vortices in the vessel and not their static
distribution, one must rotate the vessel with variable angular velocity.
Numerical calculations show that at low angular velocities one to six vortices locate symmetrically
around the cylinder axis, i. e. on a circle with its center on the axis and
radius approximately one half of the cylinder's radius. When the angular
velocity (and therefore the quantity of vortices) increases, the radius of this
circle also increases, and one more circle appears inside it, then the process
is repeated. Placing a pair of measuring electrode with appropriate
radiuses, one can observe large quantities of vortices periodically going out
of the rings.

However, these problems require more detailed analysis to which we suppose to
revert later.

\end{document}